\documentclass[prb,aps,twocolumn,superscriptaddress,longbibliography]{revtex4-1}
\usepackage{graphicx}
\usepackage{epstopdf}
\usepackage{array,multirow,amsmath,amsfonts}
\usepackage{txfonts}
\usepackage{color}
\usepackage[colorlinks,linkcolor=blue,citecolor=green]{hyperref}

\newcolumntype{Y}{>{\centering\arraybackslash}X}
\newcolumntype{C}[1]{>{\centering\let\newline\\\arraybackslash\hspace{0pt}}m{#1}}

\begin{document}
\preprint{}

\title{Strain-induced tuning of the electronic Coulomb interaction in 3$d$ transition metal oxide perovskites
 }
\author{Bongjae Kim}\email{bongjae.kim@kunsan.ac.kr}
\affiliation{University of Vienna, Faculty of Physics and Center for Computational Materials Science, Vienna, Austria}
\affiliation{Department of Physics, Kunsan National University, Gunsan, 54150, Korea}
\affiliation{Max Planck POSTECH/Hsinchu Center for Complex Phase Materials, Pohang University of Science and Technology, Pohang 37673, Korea}
\author{Peitao Liu}
\affiliation{University of Vienna, Faculty of Physics and Center for Computational Materials Science, Vienna, Austria}
\author{Jan M. Tomczak}
\affiliation{Institute of Solid State Physics, TU Wien, A-1040 Vienna, Austria}
\author{Cesare Franchini}
\affiliation{University of Vienna, Faculty of Physics and Center for Computational Materials Science, Vienna, Austria}

\date[Dated: ]{\today}

\begin{abstract}
 Epitaxial strain offers an effective route to tune the physical parameters in transition metal oxides. So far, most studies have focused on the effects of strain on the bandwidths and crystal field splitting, but recent experimental and theoretical works have shown that also the effective Coulomb interaction changes upon structural modifications. This effect is expected to be of paramount importance in current material engineering studies based on epitaxy-based material synthesization. Here, we perform constrained random phase approximation calculations for prototypical oxides with a different occupation of the $d$ shell, LaTiO$_3$ ($d^1$), LaVO$_3$ ($d^2$), and LaCrO$_3$ ($d^3$), and systematically study the evolution of the effective Coulomb interactions (Hubbard $U$ and Hund's $J$) when applying epitaxial strain. Surprisingly, we find that the response upon strain is strongly dependent on the material. For LaTiO$_3$, the interaction parameters are determined by the degree of localization of the orbitals, and \emph{grow} with increasing tensile strain. In contrast, LaCrO$_3$ shows the opposite trends: the interactions parameters \emph{shrink} upon tensile strain. This is caused by the enhanced screening due to the larger electron filling. LaVO$_3$ shows an intermediate behavior.
\end{abstract}


\maketitle

\section{Introduction}

 Transition metal oxides (TMOs) are a class of materials that are at the core of the research in
 condensed matter physics~\cite{Imada1998}. In TMOs, the electronic structure is governed by the competition between
 the local Coulomb interaction (Hubbard $U$) and the bandwidth ($W$) of the $d$-shell of the transition metal (TM)~\cite{Mott1949}. The strength of $U$ depends primarily on the spatial extent of the $d$-orbitals and on the orbital filling. It reaches its largest value for 3$d$ TMs that are most localized~\cite{Solovyev1996,Blugel2011,Vaugier2012}.
 In systems with partially-filled TM $d$-orbitals subjected to crystal-field splitting ($\Delta$) and structural distortions,
 the subtle coupling of the orbital-spin-lattice degrees of freedom gives rise to a rich variety of interesting phenomena~\cite{Imada1998,Ramirez1997,Norman2011,Cheong2007}.
 Elucidating the underlying microscopic mechanism requires command over this multitude of competing energy scales, and has been a continuous challenge for both experimental and theoretical physicists.

 The recent development of precise epitaxial growth techniques has contributed to further expanding the research in this field.
 With the help of modern synthesis techniques, target materials can be coherently strained with specific in-plane lattice parameters of various substrates~\cite{Schlom2007}. This changes the structural connectivity of the system (bond-lengths and bond-angles)~\cite{Rondinelli2011, BKim2017_1, BKim2017_2}, modifies the relative strength of the energy scales ($U$, $W$, and $\Delta$)~\cite{BKim2017_1, BKim2017_2} and leads to dramatic changes of the physical properties
 including: metal-to-insulator transition (MIT)~\cite{Rata2007,He2012,Yoshimatsu2016,Dymkowski2014,BKim2017_2,Zhong2015}, magnetic order changes~\cite{JHLee2010,JHLee2013,Hou2014,Roqueta2015,Liu2015,BKim2017_2,Liang2015}, spin-flipping~\cite{Csiszar2005,Liu2015, BKim2017_1}, enhancement of the critical temperatures~\cite{Adamo2009,Hicks2014}, \emph{etc}.
 Moreover, with proper tuning of the physical parameters by epitaxial strain, artificial engineering of novel material properties has been investigated in the field of ferroelectricity~\cite{Haeni2004}, multiferroicity~\cite{Fennie2006,Lee2010}, and superconductivity~\cite{Hansmann2009,Stewart2011,Ivashko2018}.

 Typically, the parameters that are considered to be tunable by epitaxial strain are $W$ and $\Delta$.
 Coherent strain induces changes in the bond-lengths and the rotation/tilting angles of TM-O polyhedra, which affects the electron hopping between sites and in turn modifies the size of $W$. The energy levels of the $d$-orbitals can also be varied upon strain, for instance lifting the degeneracies of $t_{2g}$ or $e_{g}$ levels, and the corresponding changes in $\Delta$ could lead to the onset of charge or orbital ordered phases. In a recent extensive study of the role of epitaxial strain on $d^1$ and $d^2$ TMO perovskites, LaTiO$_3$  and LaVO$_3$, it was shown that a MIT can be selectively induced upon strain by the modifications of $W$ and $\Delta$~\cite{Sclauzero2016}. In the case of the Mott insulator LaTiO$_3$ (LTO), epitaxial strain can enhance the electronic hopping by changing $\Delta$, explaining the experimentally observed metallic behavior under compressive strain~\cite{Dymkowski2014,He2012}. In contrast to LTO, LaVO$_3$ (LVO) shows a more robust insulating character upon strain, which suggests that the metallicity found in the LVO film on top of SrTiO$_3$ substrate might arise from interfacial effects, rather than from intrinsic changes~\cite{Sclauzero2015,Hotta2007,He2012}.

 Despite intensive studies, the role of strain on the strength of the effective Coulomb interaction $U$ has been largely overlooked.
 $U$ is typically treated as a constant, insensitive to strain. This choice is justified by the argument that the range of coherent epitaxy strain is too small ($<$ 5~\%) to induce substantial changes on $U$, and that the formal occupation of the $d$-orbital remains unchanged. Consequently, in first principles calculations based on density functional theory (DFT) supplemented by an Hubbard-type Coulomb parameter (DFT+$U$) as well as in dynamical mean-field theory (DMFT) calculations, it is  a common practice to keep the value of $U$ fixed over the entire range of the strains~\cite{Dymkowski2014,Sclauzero2015}. However, it has been reported that even small structural variations can have large influences on $U$~\cite{Tomczak2009,Tomczak2010,Nilsson2017,Kunes2008,Leonov2015}. This is, \emph{e. g.}, the case for MnO, where the effective $U$ was found to increase upon pressure as a consequence of the changes of the TM structural environment~\cite{Tomczak2010}. In Bi$_2$CuO$_4$, on the other hand, the Hubbard $U$ shrinks under pressure, potentially triggering an insulator-to-metal transition~\cite{DiSante2017}. These results are indicative of a delicate interplay between the local structure and the screening properties, which unavoidably influences the strength of $U$~\cite{Tomczak2009,Nilsson2017,DiSante2017}. In a recent study on iridates~\cite{Nichols2013}, systematic shifts of optical peaks have been observed in coherently strained samples, suggestive of a direct observation of changes in the effective $U$ upon strain. Also for cuprate (La$_2$CuO$_4$) thin films, substrate-induced strain was recently shown to significantly tune effective interactions, such as the Hubbard $U$ and the magnetic exchange coupling\cite{Ivashko2018}.

 To address this issue, in this paper we aim to study the role of the epitaxial strain on $U$ by computing effective low-energy interactions within the constrained random phase approximation (cRPA) at different strain levels for a representative set of 3$d$ TMO perovskites with different orbital-occupancies: LTO ($t_{2g}^1$), LVO ($t_{2g}^2$), and LaCrO$_3$ (LCO, $t_{2g}^3$)~\cite{He2012CF}. We show that the electronic Coulomb interactions are strongly material-dependent, in particular in its response to epitaxial strain. This is caused by the delicate competition between the degree of localization of the correlated orbitals and the screening arising from the $d$-$p$ hybridization.

\section{Methods}

 We performed {\it ab initio} electronic structure calculations using the projector augmented wave method employing the Vienna {\it ab initio} simulation package (VASP) \cite{Kresse1993,Kresse1996}. For the exchange-correlation functional, we adopted the generalized gradient approximation by Perdew-Burke-Ernzerhof (PBE) and a plane-wave cutoff of 400 eV was used~\cite{Perdew1996}. For the bulk systems, we used the experimental unit cells (20 atoms) which contains a $a^-a^-c^+$ type tilting~\cite{Cwik2003,Bordet1993,Li2002} . To simulate the epitaxial strain, we first fully relaxed the bulk unit cell to identify the equilibrium volume and the equilibrium lattice parameters $a$, $b$, and $c$, and set $a_0=\sqrt{(a^2+b^2)}$ as the 0~\% strain limit. 
 Tensile and compressive strains (up to $\pm$4~\%) are obtained by performing full structural relaxation within the tetragonal symmetry for different values of the in-plane lattice parameter with an accuracy of $10^{-3}$~ eV/\AA.
 Monkhorst-Pack $k$-meshes of 6$\times$6$\times$4 were used.

 To quantify the screened Coulomb interaction parameters, we adopted the cRPA~\cite{Aryasetiawan2004}.
 The central idea of cRPA is to exclude all the screening channels within the target correlated subspace (usually $d$ orbitals in TMOs)
 ${P}^{c}$ from the total polarizability  ${P}$
\begin{equation}
{P}^{r} = {P} - {P}^{c}.
\label{eq:1}
\end{equation}%
Then the partially screened Coulomb interaction kernel $\mathbf{U}$ can be obtained by solving the following Bethe-Salpeter equation
\begin{equation}
\mathbf{U}^{-1}= [\mathbf{U}^{bare}]^{-1} - {P}^{r},
\label{eq:2}
\end{equation}%
 where $\mathbf{U}^{bare}$ are bare (unscreened) interactions. In the present work, the correlated subspace is chosen as the $t_{2g}$ orbitals of the TM, which are constructed by means of maximally localized Wannier functions obtained by the Wannier90 code~\cite{Marzari1997,Mostofi2008,Franchini2012}. The detailed procedure of our cRPA method can be found in Ref.~\onlinecite{Nomura2012}.

To evaluate the strain-dependent evolution of the interactions, we have tested two different setups, $t_{2g}$/$t_{2g}$ and $t_{2g}$/$t_{2g}$-$p$,  to explicitly demonstrate that the underlying physics is not scheme-dependent~\cite{Vaugier2012,Panda2017}. The difference between the two models lies in the way the local orbital basis are obtained. In the $t_{2g}$/$t_{2g}$ model Wannier functions are constructed for TM-$t_{2g}$ only, whereas in the $t_{2g}$/$t_{2g}$-$p$  model not only TM-$t_{2g}$ but also O-$p$ Wannier functions are constructed.
However, in both models the interaction parameters are obtained for the TM-$t_{2g}$ subspace, which govern the low-energy physics.
The resulting Coulomb $U$ and the Hund's coupling parameters $J$ are obtained by averaging the $\mathbf{U}_{ijij}$ and $\mathbf{U}_{ijji}(i\neq j)$ matrix elements as calculated from
\begin{equation}
\mathbf{U}_{ijkl} = \lim_{\omega \rightarrow 0} \iint d^3r d^3r^\prime w_i^{*}(\mathbf{r}) w_k^{*}(\mathbf{r^\prime}) %
\mathbf{U} (\mathbf{r},\mathbf{r^\prime},\omega) w_j(\mathbf{r}) w_l(\mathbf{r^\prime}),
\label{eq:3}
\end{equation}%
where $w(\mathbf{r})$ refers to $t_{2g}$-like Wannier functions. Detailed computational information is described in Ref.~\cite{Kaltak2015}. Other effects such as the frequency dependence are not discussed in the present study, though technically it is possible to include these effects~\cite{Aryasetiawan2004,Tomczak2017}.

\section{Results and discussions}

\subsection{Bulk phases}

\begin{figure}[t]
\begin{center}
\includegraphics[width=0.99\columnwidth,clip=true]{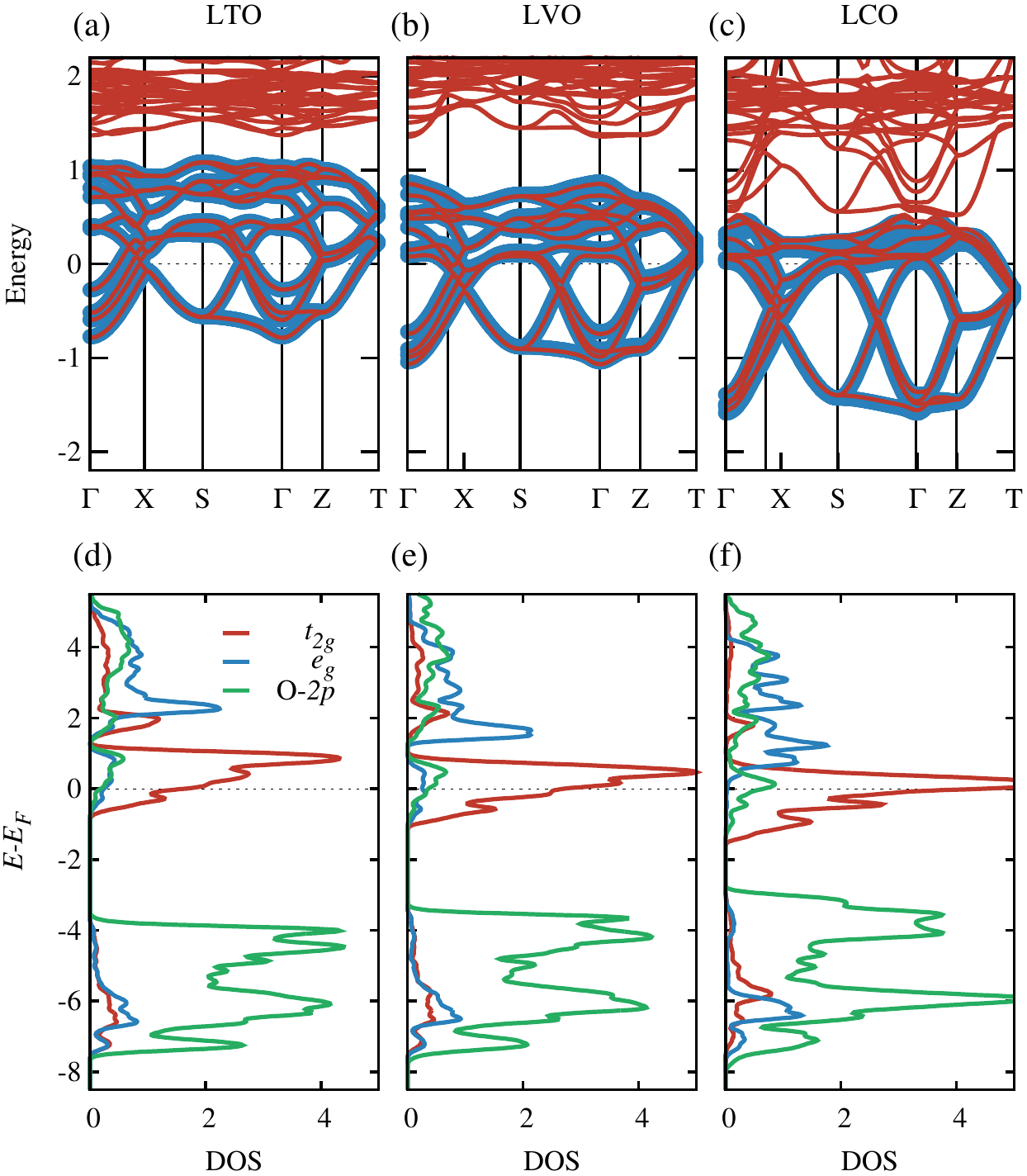}
\end{center}
\caption{
(a)-(c): Nonmagnetic band structures of LTO, LVO, and LCO together with their Wannier-projected bands. {\it Ab initio} bands and Wannier-projected bands are denoted with thin (red) and thick (blue) lines, respectively. Fermi level has been aligned to zero.
 (d)-(f): Partial density of states (DOS) for three systems at their 0\% strain cases. $e_g$ and $t_{2g}$ orbitals (blue and red) of TM ions are separated by the octahedral symmetry, and O-$p$ orbitals are located well below the Fermi level. The energy separation between $t_{2g}$ and $p$ orbitals is decreasing from LTO to LCO.
}
\label{fig1}
\end{figure}

 Before discussing the effect of strain, we examine the electronic and screening properties of the bulk compounds LTO ($t_{2g}^1$), LVO ($t_{2g}^2$), and LCO ($t_{2g}^3$).
 In general, it is expected that the localization of the $d$ orbitals should increase with orbital occupation along the same TM row of the periodic table, and the corresponding contraction of the correlated $d$ space with atomic number should lead to a reduction of  the hybridization between TM-$d$ and O-$p$ orbitals~\cite{Vaugier2012}. The other hybridization channels are governed by the energy separation between O-$p$-$d$ and $d$-$d$ orbitals, which is also sensitive to the atomic number.
 All three compounds under scrutiny are antiferromagnetic insulators with GdFeO$_3$-type (GFO) tilting, which is common for perovskite oxides. In Fig.~\ref{fig1}, nonmagnetic band structures and the corresponding density of states (DOS) are shown together with the Wannier-interpolated bands for the $t_{2g}$ states. The $t_{2g}$ bands develop in an energy window of about $\pm$1~eV around the Fermi level, and are separated from the empty TM-$e_g$ and filled O-$p$ bands (with the exception of LCO) as well as from the underlying occupied O-$p$ states, though $t_{2g}$-O-$p$ hybridization takes place near the Fermi energy (See Fig.~\ref{fig1} (d)-(f)).
 The $t_{2g}$ bands are progressively pushed down in energy with increasing electron filling and the $t_{2g}$-$e_g$ and $t_{2g}$-O-$p$ gaps are continuously reduced. In LCO the $t_{2g}$ manifold starts to mix with the unoccupied states above. This trend is the key to understand the different screening properties of the system.
 This minimal interpretation of the electronic structure suggests that the low-energy physics of the systems is mainly determined by the $t_{2g}$ states, and the effects of entanglement and hybridization with other states are not as crucial as in heavier $4d$- and $5d$-TMOs~\cite{Vaugier2012,Ergonenc2018}.

\begin{table}[b]
\centering
\caption{Computed $U$ and $J$ parameters for both $t_{2g}/t_{2g}$ and $t_{2g}/t_{2g}$-$p$ models for bulk LTO, LVO, and LCO systems. Units are in eV.}
\begin{tabular}{C{0.7cm}C{0.8cm}C{0.8cm}C{0.8cm}C{0.8cm}|C{0.8cm}C{0.8cm}C{0.8cm}C{0.8cm}}
\hline\hline
                          &\multicolumn{4}{c}{$t_{2g}/t_{2g}$}&\multicolumn{4}{c}{$t_{2g}/t_{2g}$-$p$}      \\
                          &       $U$     &  $U^{bare}$ &  $J$&  $J^{bare}$  &       $U$     & $U^{bare}$&   $J$&  $J^{bare}$  \\
\hline
      LTO                 & 2.49          &  11.78      & 0.35&  0.35  &  3.57         & 13.34     &0.45  &  0.50        \\
      LVO                 & 2.92          &  14.61      & 0.43&  0.50  &  3.69         & 16.27     &0.53  &  0.60        \\
      LCO                 & 2.40          &  15.79      & 0.43&  0.52  &  3.21         & 18.09     &0.55  &  0.65        \\
\hline\hline
\end{tabular}
\label{bulk}
\end{table}

 The calculated unscreened and screened local interactions obtained by cRPA using, both, the $t_{2g}$/$t_{2g}$ and $t_{2g}$/$t_{2g}$-$p$ models are compiled in Table~\ref{bulk}. Both models deliver essentially the same picture. The only quantitative difference is the enhancement of the $U$ and -- to a lesser extent -- $J$ values within the $t_{2g}$/$t_{2g}$-$p$  model, originating from the inclusion of the O-$p$ states in the Wannier projection which leads to a higher localization of the $t_{2g}$ orbitals.

\begin{figure}[h]
\begin{center}
\includegraphics[width=0.99\columnwidth,clip=true]{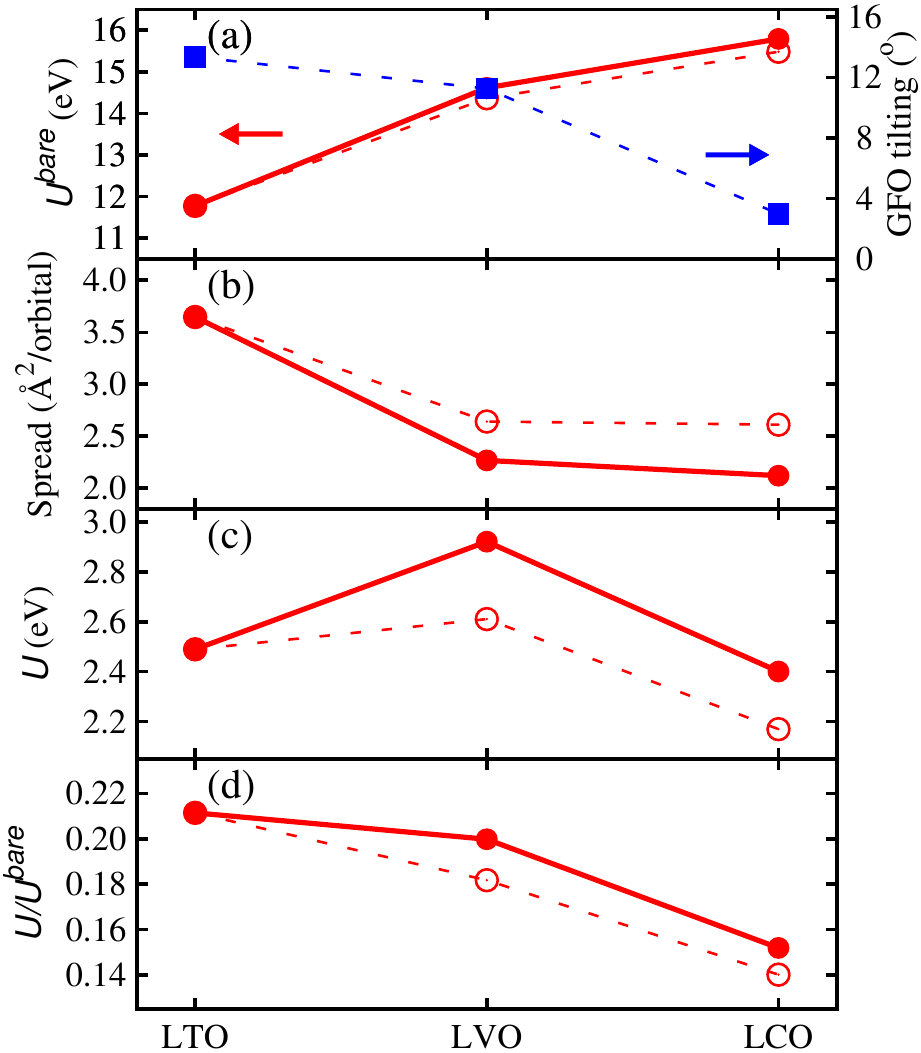}
\end{center}
\caption{
Evolution of the cRPA Hubbard parameters ($U$, $U^{bare}$ and $U^{bare}/U$, within the $t_{2g}$/$t_{2g}$ model) and the spread of the Wannier function are shown in red solid lines with filled circles. Average GFO tilting  $\alpha$ (blue dotted line with filled squares) are also shown. Experimental values~\cite{Cwik2003, Bordet1993, Li2002}) were employed along the series LTO, LVO and LCO for the structurally distorted bulk phases. Results for LVO and LCO in the LTO structures are shown with red dashed lines with open circles.
}
\label{fig2}
\end{figure}

 Based on the trends of the $t_{2g}$/$t_{2g}$ interaction parameters along the series (see Fig.~\ref{fig2}) we can draw the following physical picture. We first notice that the unscreened interaction ${U^{bare}}$, which measures the degree of electron localization without including screening effects, increases by as much as 4.0 eV ($\approx$ 34\%) from LTO to LCO. This behaviour can be either attributed to the decrease of GFO distortions, or to purely electronic effects associated with the contraction of the $d$ orbitals with increasing electron filling~\cite{Vaugier2012}.

 From the structural point of view, from LTO to LCO the TM atomic radius decreases from 0.67 (Ti) to 0.615 (Cr), which causes a continuous shrinking of the volume and a linear increase of the tolerance factor from 0.948 (LTO) to 0.975 (LCO)~\cite{He2012CF}. Perovskites with a lower tolerance factor are more inclined to structural distortions, in this case the GFO distortions, that can be quantified by the average tilting angle $\alpha=\frac{1}{2}[180-\theta$] ($\theta=\widehat{{TM}-{\rm O}-{TM}}$) of the in-plane ($\alpha_{IP}$) and out-of-plane ($\alpha_{OP}$) tilting angles. We can notice that the GFO distortion is largest for LTO (see Fig.~\ref{fig2}(a)), and, accordingly, tilting and rotation of the TM-O octahedra induces hybridization and broadening of the $t_{2g}$ bands, as confirmed by LTO having the largest spread (3.6 \AA$^2$/orbital). GFO-type distortion progressively decreases from LTO via LVO to LCO, and the average tilting angle is largely decreased from 13$^\circ$ (LTO) to 3$^\circ$ (LCO).

 To separate out this strong structural effects from changes in the electron occupation, we calculated the various parameters of LVO and LCO using the LTO structure, which are denoted with dashed lines in Fig.~\ref{fig2}. The increase of ${U^{bare}}$ upon electron filling should be attributed to the enhanced localization of the $t_{2g}$ orbitals, which can be quantified from the spread of the Wannier functions. The spread provides a measure of the degree of localization of the orbitals and is correlated with the magnitude of the hopping integrals~\cite{Tomczak2010}: the larger is the spread, the larger are the hopping integrals. As shown in Fig.~\ref{fig2}(b), the overall spread decreases upon electron occupations, which indicates the increase of the spatial localization of $t_{2g}$ orbitals from LTO to LCO. This is directly reflected in the behaviors of $U^{bare}$. We find that the structural distortion only plays a minor role in this trend (see Fig.~\ref{fig2} (a) and (b)).

 Interestingly, the Hubbard $U$ does not follow the trend of $U^{bare}$ (Fig.~\ref{fig2} (c)). Indeed, $U$ is {\it non-monotonic} along the series: Rising from 2.49 to 2.92~eV from LTO to LVO, it drops back to 2.40~eV in LCO. This reversal of trend is indicative of a substantial increase of the effectiveness of screening along the series. The ratio $U/U^{bare}$ decreases continuously from 0.21 for LTO to to 0.15 for LCO. This is predominantly due to the reduction of the $O_p$-$t_{2g}$ and $e_g$-$t_{2g}$ gaps, which intensifies the corresponding screening channels. Screening effects are particularly strong for LCO due to the complete closing of the $e_g$-$t_{2g}$ above the Fermi level as well as the intraband gap within the $e_g$ manifold (see Fig.~\ref{fig1}) and causes a huge shrinking of the local interaction from 15.79 eV ($U^{bare}$) to 2.40 eV ($U$). The competition of the Wannier localization and the screening decides the screened behavior of the interaction parameters, while purely structural effects are of subleading importance.

 In conclusion, our results convey the following message: the increasing orbital localization from LTO to LCO, mostly due to electronic effects and responsible for the relatively large values of $U^{bare}$, is counterbalanced by the enhancement of screening effects (in particular for LCO), which, in the bulk phases, ultimately leads to rather similar values of $U$ for the whole perovskites series. The structural effects (GFO tilting) have only a marginal role when compared to the effects of the electronic occupations. Similar conclusions were achieved by Vaugier \emph{et. al.} for the  $t^1_{2g}$ to $t^3_{2g}$ perovskite family Sr$M$O$_3$ ($M$ = V, Cr, Mn) albeit the structural effects were not considered in their study~\cite{Vaugier2012}.

\subsection{Effect of epitaxially strain}

\begin{figure}[t]
\begin{center}
\includegraphics[width=0.99\columnwidth,clip=true]{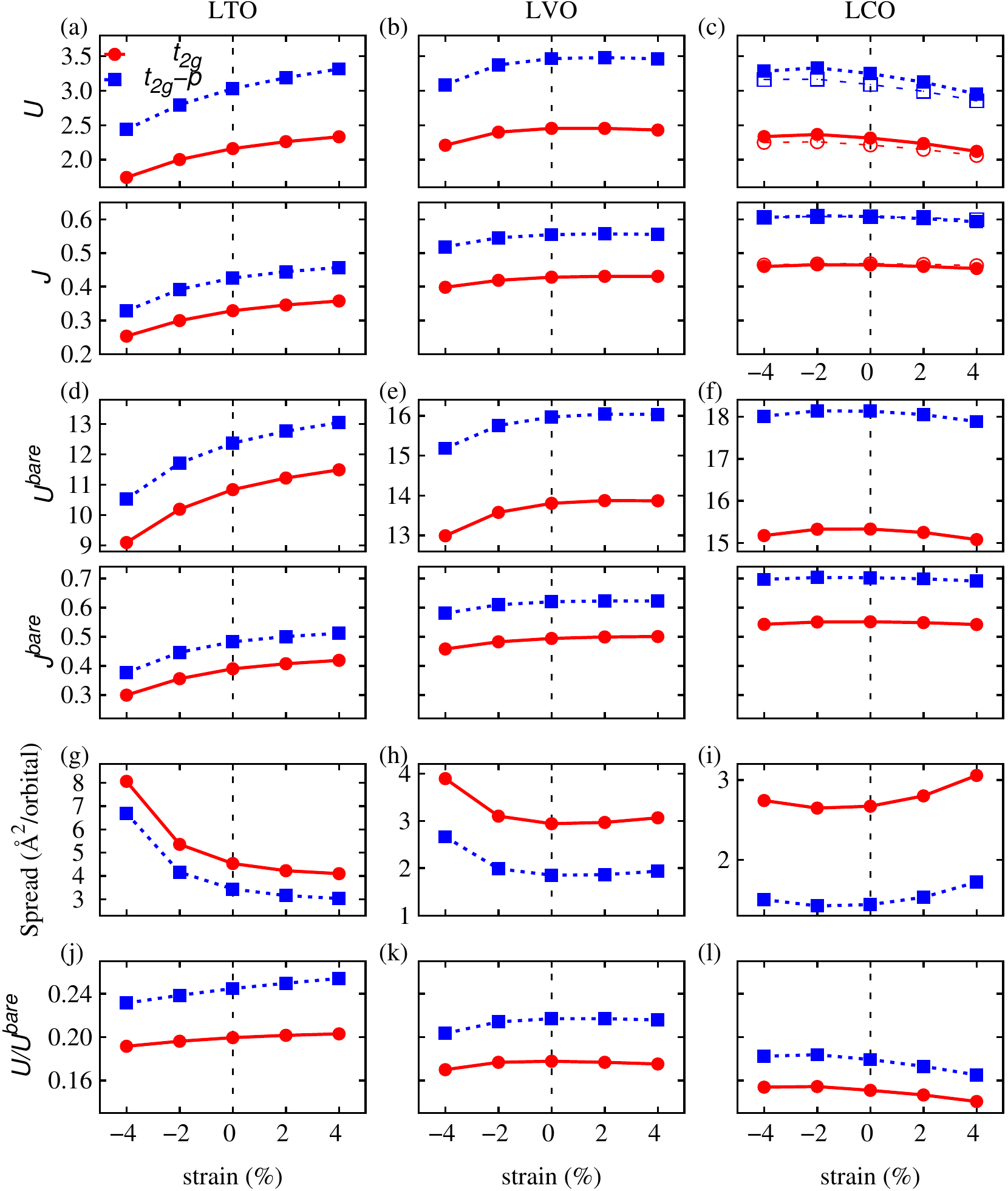}
\end{center}
\caption{
 Effective interactions $U$ and $J$ (a-c),  $U^{bare}$ and  $J^{bare}$ (d-f) for LTO, LVO and LCO
 employing both the $t_{2g}/t_{2g}$ (red) and $t_{2g}/t_{2g}$-$p$ (blue) setup as a function of epitaxial substrate strain.
  The $U$ and $J$ values of LCO obtained by using the LTO structures are shown with dashed lines in (c).
 (g-i) Spread of the Wannier functions and (j-l) $U/U^{bare}$.
}
\label{fig3}
\end{figure}

 We now address the role of epitaxial strain on the electronic interactions. Fig.~\ref{fig3} shows the calculated values of $U$, $J$, $U^{bare}$ and $J^{bare}$ as a function of epitaxial strain for the three compounds assuming a tetragonal crystal symmetry.

 First, we note that the values of $U$ and $J$ in the zero strain limit are slightly lower than the corresponding bulk reference values listed in Tab.~\ref{bulk}, due to the different crystal symmetries (orthormbic/monoclinic vs. tetragonal) which modify the degree of GFO distortion. As expected, the unscreened interactions $U^{bare}$ at zero strain increase from LTO to LCO due to higher electron fillings. At variance with the bulk case, the spread of the Wannier function \emph{decreases} monotonically along the series suggesting that in the zero strain tetragonal phases electron filling should be the dominant factor that increases orbital localization from LTO to LCO. Note that when the structures are constrained to the LTO geometry, the Wannier functions show behaviors similar to the bulk case (See open circle in Fig.~\ref{fig2} (b))

 The data clearly show that $U$ and $J$ are significantly altered by strain, but, interestingly, the changes are strongly system dependent, as elaborated below. Both models adopted for the cRPA calculations convey a qualitatively similar picture. Therefore, for the sake of clarity we limit our discussions to the $t_{2g}/t_{2g}$ model; similar conclusions are valid for the $t_{2g}/t_{2g}$-$p$ setup.

 In LTO $U$ increases monotonically by $\approx$ 34\% in the entire range of strains from -4~\% to +4~\%, whereas
 for LVO and LCO the strain-induced modifications are much attenuated, $\approx$ 12~\%, and follow different trends.
 In LVO [Fig.~\ref{fig3} (b)] $U$ increases from 2.21 to 2.45 eV from -4~\% to 0~\% and then remains pretty much constant in the tensile strain regime. In contrast, for LCO [Fig.~\ref{fig3} (c)] $U$ decreases very smoothly from the compressive to the  tensile regime. Similar trends are observed for $J$ and for the corresponding bare interaction parameters, with the exclusion of $U^{bare}$ and $J^{bare}$ in LCO which are rather insensitive to strain and remain essentially unchanged in the whole strain domain.

\begin{figure}[t]
\begin{center}
\includegraphics[width=0.99\columnwidth,clip=true]{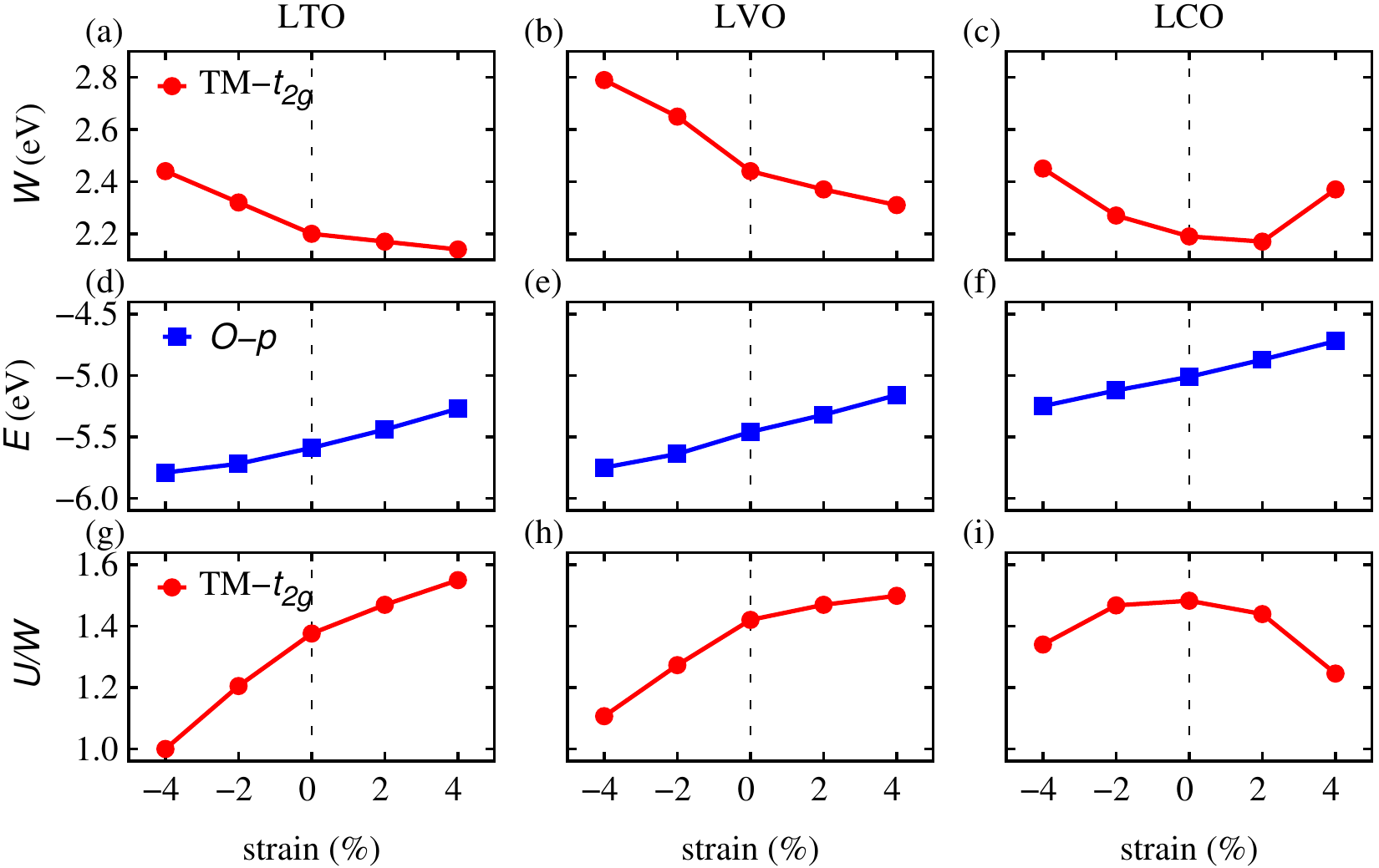}
\end{center}
\caption{
 Bandwidth of $t_{2g}$ orbitals (a-c), center of mass $E$ of the O-$p$ bands with respect to the Fermi level (d-f), and $U/W$ ratio LTO, LVO, and LCO as a function of epitaxial strain (g-i).
}
\label{fig4}
\end{figure}

\begin{figure}[t]
\begin{center}
\includegraphics[width=0.99\columnwidth,clip=true]{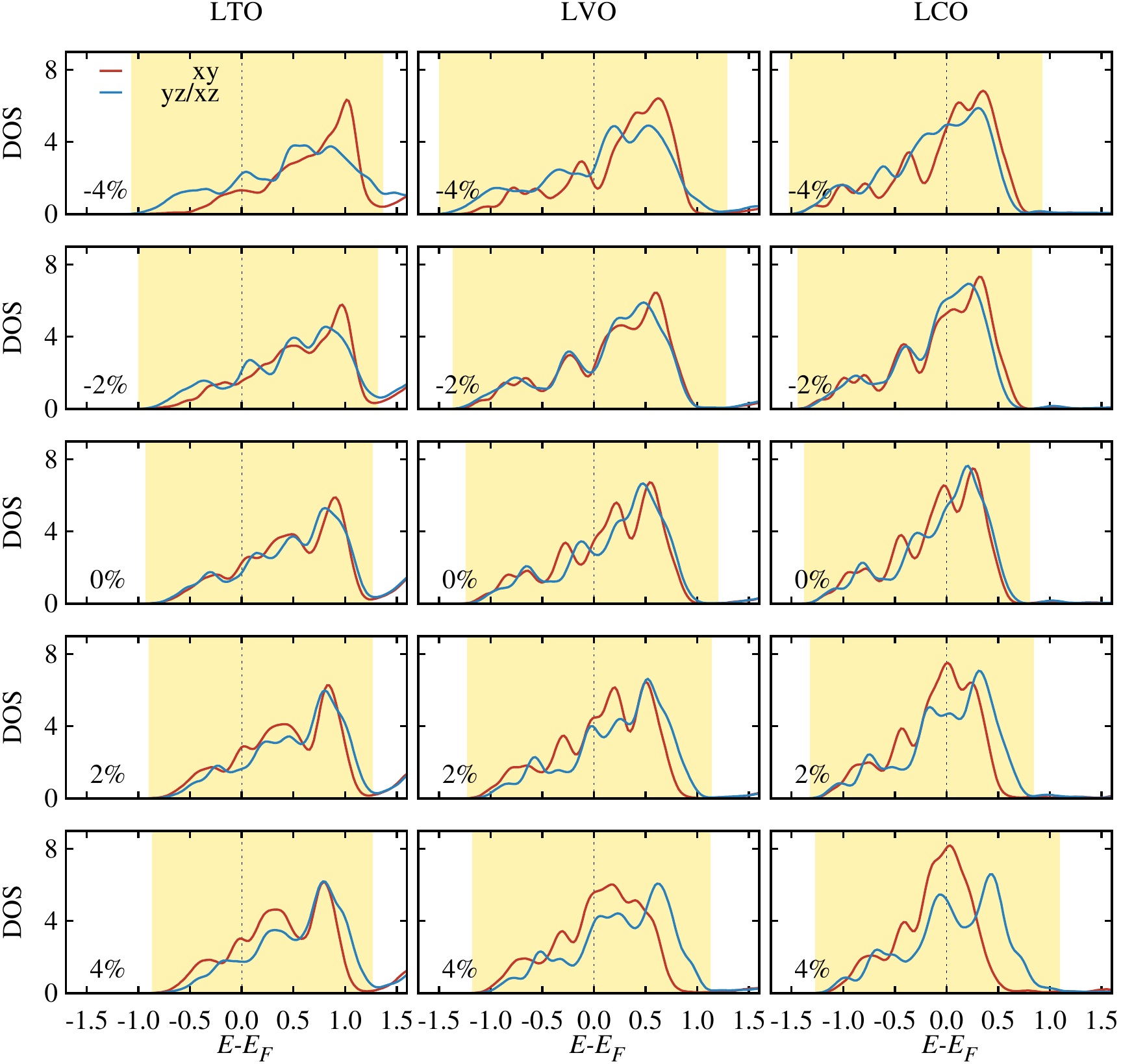}
\end{center}
\caption{ Orbital-resolved DOS for LTO, LVO, and LCO as a function of epitaxial strain. Shaded (yellow) region denotes the width of $t_{2g}$ orbitals, which is defined as bandwidth, $W$.
}
\label{fig5}
\end{figure}

 To understand the origin of these behaviours it is necessary to inspect how localization effects, hybridization and screening are altered by strain in the three materials, and how these effects are correlated with the underlying structural distortions. We first focus on the correlation between the spread and $U^{bare}$ to exclude the influence of screening effects. In LTO the spread decreases rapidly upon tensile strain from -4\% to 0\%, in accordance with the fast enhancement of $U^{bare}$  [see Fig.~\ref{fig3} (d)] and then continues to decrease monotonically in the tensile-strain region. This behavior is well correlated with the evolution of the bandwidth $W$. LVO follows a similar trend but the overall shrinking of the spread from -4\% to +4\% is reduced by 50\% with respect to LTO. The anomalous decrease of $U$ observed for LCO [Fig.~\ref{fig3} (c)] is reflected in a different change of the spread upon strain, which varies only little going from  -4\% to 0\% and then increases by about $\approx$ 15\% for tensile strain. The reduced localization in LCO for positive strain can be connected with the increase of $W$ for positive strain, see Fig.~\ref{fig4}(c) and right panels of Fig.~\ref{fig5}. The decrease of the spread and of the bandwidth $W$ is associated with larger orbital localization, which explains the increasing  $U^{bare}$ for LTO and LVO upon strain, as well as the anomalous decrease of  $U^{bare}$ for LCO for positive strain.

 Now we investigate the screening effects in the systems. For the case of hydrostatic pressure, the local environment of the TM ion changes in an isotropic manner, and leads to the increase of the hybridization and bandwidth. However, the behavior of $U$ is found to be not only dependent on the delocalization of the Wannier functions~\cite{Tomczak2009}, but also on the significant weakening of the screening~\cite{Tomczak2010}. We can expect that similar physics develops for the epitaxial strain case, where the structural behaviors in general show the opposite trends for in-plane and out-of-plane directions. To evaluate the screening effect, we plotted $U/U^{bare}$ for all three compounds upon epitaxial strain in Fig.~\ref{fig3}(j)-(l). As manifested by the decrease of $U/U^{bare}$ values, the screening becomes stronger as the occupation of the $t_{2g}$ orbital increases from LTO ($t_{2g}^1$) via LVO ($t_{2g}^2$) to LCO ($t_{2g}^3$). The enhanced screening for larger occupation can be understood from the overlap of bands between O-$p$ and TM-$t_{2g}$ orbitals. As clearly seen in Fig.~\ref{fig1} (d)-(f), for increasing occupancy from LTO via LVO to LCO, TM-$t_{2g}$ levels move to lower energy while the O-$p$ states shift upwards to the Fermi level. In Fig.~\ref{fig4}(d)-(f), we plotted the center of mass of the O-$p$ level with respect to the Fermi level, where the relative position of O-$p$ DOS moves up from LTO to LCO. This promotes enhanced screening from the $t_{2g}$-$p$ channel as the occupation increases~\cite{Vaugier2012}.

 For LTO, screening processes do not change the overall response of $U^{bare}$ upon strain: also the screened parameter, $U$, increases upon tensile strain. For LVO and LCO, the variation of the spread is, however, much smaller than in the case of LTO, which explains why the response of the bare parameters to the epitaxial strain is much weaker: changes of $U^{bare}$ are $\sim$0.9~eV and 0.1~eV for LVO and LCO, respectively. Interestingly, the modification of the screened Coulomb parameters ($\sim$0.2~eV) is even larger than the bare one for LCO, which demonstrates that the {\it decrease} in $U$ for LCO has a different origin as compared to LTO, where the localization of the orbitals drives the {\it increase} of $U$.

 Thus, the intensity of screening effects appears to be a key factor in determining the different responses to strain in the three different materials. As tensile strain is applied in LTO (see Fig.~\ref{fig3}(j)-(l)), screening effects become weaker as seen from the monotonic increase of $U/U^{bare}$. But for LCO, the screening becomes stronger as evident from the attenuation of $U/U^{bare}$, which fits well with the decreasing behavior of $U$ for LCO. This clearly demonstrates that for LCO the strain-dependency of $U$ is dominated by the screening, not by the Wannier localization. As we noted before, for LCO the decreasing behavior of $U$ in spite of an almost unchanging $U^{bare}$ upon strain indicates the prominent role of the screening. We can see that the relative position of O-$p$ state increases upon strain for all three cases Fig.~\ref{fig4}(d)-(f) but screening from $t_{2g}$-$p$ channel seems to be more effective when the occupation is larger.

 We find that the overall degree of electronic correlations, quantified by $U/W$, increases for LTO and LVO upon tensile strain (Fig.~\ref{fig4}(g) and (h)). Unexpectedly, we found nonmonotonic behaviors of $U/W$ in LCO (see Fig.~\ref{fig4}(i)), which is due to the crystal field splitting $\Delta$ between $xy$ and $yz/zx$ orbitals, which is much larger for LCO than LTO, especially for the tensile limit (see Fig.~\ref{fig5}). For the tensile strain case in LCO, the larger $\Delta$ extends the energy range of $t_{2g}$ orbitals and increases the value of $W$. In combination with a reduced $U$, the overall $U/W$ shows decreasing behaviors for LCO upon tensile strain.

\begin{figure}[t]
\begin{center}
\includegraphics[width=0.99\columnwidth,clip=true]{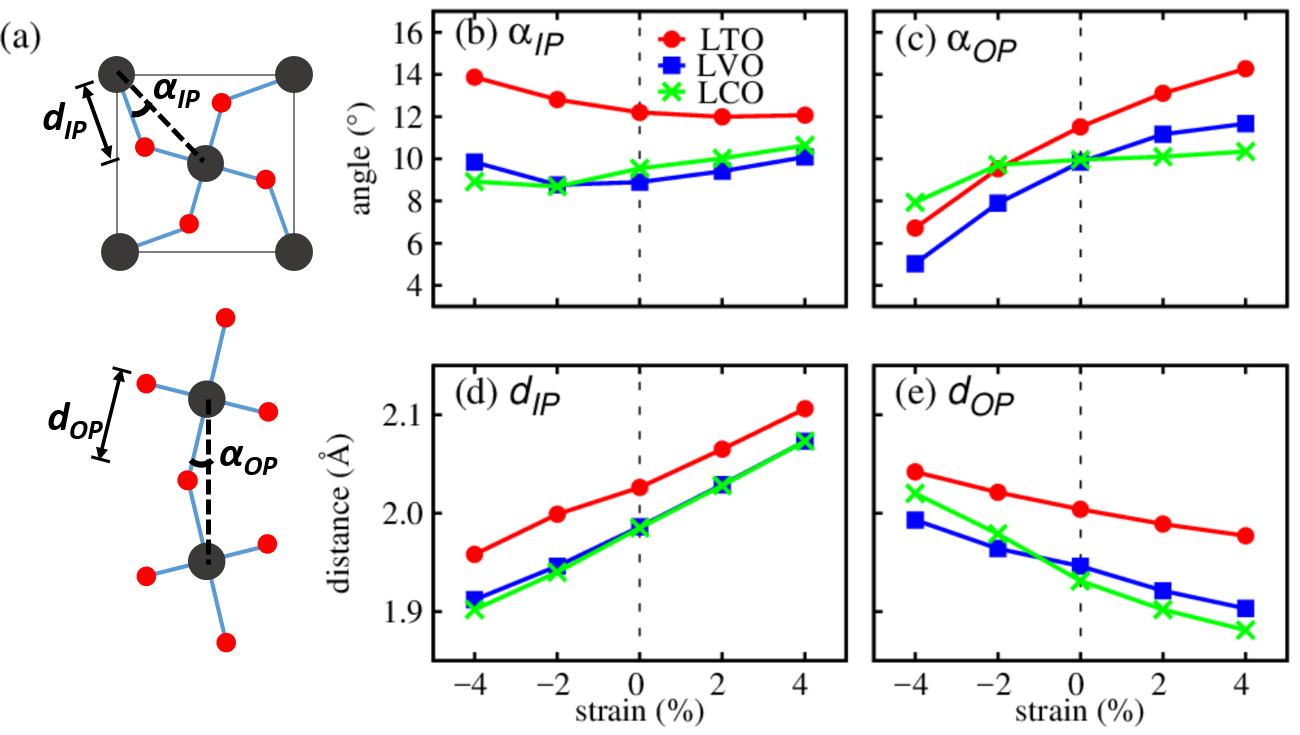}
\end{center}
\caption{ (a) The schematic diagram showing IP and OP bond angles and bond lengths. Evolution of (b) $\alpha_{IP}$, (c) $\alpha_{OP}$, (d) $d_{IP}$ and (e) $d_{OP}$  upon epitaxial strain.
}
\label{fig6}
\end{figure}

 Now we briefly discuss strain-induced structural effects. In general, for ABO$_3$-type perovskite, the response of the bond angles ($\alpha_{IP}$ and $\alpha_{OP}$) and bond lengths ($d_{IP}$ and $d_{OP}$) upon epitaxial strain are different for in-plane (IP) and out-of-plane (OP) directions (see Fig.~\ref{fig6}(a)). As the tensile strain is applied, $\alpha_{IP}$ ($\alpha_{OP}$) is expected to decrease (increase) and $d_{IP}$ ($d_{OP}$) to increase (decrease). IP strain effects are compensated by the changes in the OP connectivity as found for LTO (Fig.~\ref{fig6}(b)-(e)). In Fig.~\ref{fig4}(b)-(e), we clearly see that the variation in $\alpha_{IP}$ is quite small ($\leq 2^{\circ}$) while $d_{IP}$ increases progressively upon tensile strain. This is different for the apical direction, where the variation in angle is larger while the $d_{OP}$ decreases moderately. Considering that the bond length between atoms has more direct effects than the bond angles to the hopping integrals~\cite{Sclauzero2016}, we can conclude that structural changes in the IP direction are more important than the OP ones. This is corroborated by the fact that the overall bandwidth $W$ decreases for all three cases (See Fig.~\ref{fig4}(a)-(c)).

 Noteworthy is that we observe an unexpected upturn of the $\alpha_{IP}$ for LVO and LCO (Fig.~\ref{fig4}(b)), which cannot be explained within a rigid MO$_6$ octahedron model. This counterintuitive behavior was previously reported from other {\it ab initio} studies~\cite{Dymkowski2014,Sclauzero2015}, but further confirmation from the experiment is needed. To check weather the abnormal response of $\alpha_{IP}$ is related to the decreasing $U$ for LCO, we performed cRPA calculations of LCO with structural parameters of LTO for all strain ranges (exchanging Ti with Cr for LTO-relaxed structures). The resulting $U$ and $J$ parameters follow the trends of LCO as shown in Fig.~\ref{fig3} (c), which excludes structural effect as the origin of the different response for LTO and LCO.

 As a final note, we want to briefly discuss the importance of strain on MIT. Bulk LTO is a Mott-type insulator, but shows metallic behavior in the form of compressively strained films~\cite{He2012}. Whether the origin of the observed metallicity is due to strain or interfacial effects needs further discussions. In a recent study by Dymkowski and Ederer~\cite{Dymkowski2014}, the metallicity observed for compressive strain is attributed to the enhanced hopping between two lowest $t_{2g}$ levels, $yz$ and $xz$, originated from the crystal field effect. Accordingly, the critical $U$ parameter ($U^c$) for the onset of the MIT strongly depends on the strain such that $U^c$ increases upon compressive strain (See also Di Sante \emph{et al.} for a similar example in the case of hydrostatic pressure~\cite{DiSante2017}). Based on our findings of a decreasing $U$ upon compressive strain in LTO, we claim that the metallic behavior found in the experiment does not arise from the interfacial effects but rather from the epitaxial strain~\cite{He2012}. For LVO, it was shown that the specific type of interface can decide the insulating or metallic phase of the system~\cite{Hotta2007}, and calculations showed that LVO has a robust insulating character upon strain~\cite{Sclauzero2015}. Considering that the modification of $U$ for LVO is modest compared to LTO as shown in our study, the metallic behaviors found for LVO systems on substrates should have an interfacial origin~\cite{He2012}. As LCO is a very robust insulator with a large gap~\cite{Arima1993,Maiti1996}, we think it unlikely that a metallic phase could be reached soley by means of epitaxial strain.

\section{Conclusions}

 In summary, we have systematically studied the role of the epitaxial strain onto effective electronic interactions. By taking representative TMO perovskites LTO, LVO, and LCO as examples, we have shown that the Hubbard $U$, conventionally treated to be a fixed parameter insensitive to structural changes, is strongly dependent on epitaxial strain. Interestingly, as tensile strain is applied, LTO and LCO show different behaviors: $U$ {\it increases} for LTO due to the enhanced localization of the $t_{2g}$ orbitals, while for LCO $U$ {\it decreases} due to the strong screening arising from the enhanced $d$-$p$ band overlaps. We hope that experimental investigations on 3$d$ systems employing spectroscopic techniques such as inverse photoemission spectroscopy will confirm our findings similarly to what has been recently done for 5$d$ oxides~\cite{Nichols2013}.

 Our study demonstrates the importance of the interplay between structural and electronic degrees of freedom in TMOs, where the dominant physical parameters are often competing within narrow energy scales. For the design of {\it realistic} functional materials based on heterostrucures and epitaxial films, we assert that the modification of Coulomb $U$ should be considered, especially when theoretically assessing the optical properties~\cite{PLiu2018,MKim2018}.

\section{Acknowledgements}
We thank C. Ederer, M. Kaltak, G. Kresse, S. Panda and S. Biermann for fruitful discussions. This work was supported by the joint Austrian
Science Fund (FWF) and Indian Department of Science and Technology (DST) project INDOX (I1490-N19) and by the FWF-SFB ViCoM (Grant No. F41). B.K. acknowledges support by the National Research Foundation of Korea (NRF) under the project No. 2016K1A4A4A01922028. Computing time at the Vienna Scientific Cluster is greatly acknowledged.

\bibliographystyle{apsrev4-1}
\bibliography{bibfile}



\end{document}